# ON THE EXPERIMENTAL SUBSTANTIATION OF ANISOTROPY OF INERTIAL MASS OF BODY IN THE EARTH' GRAVITATIONAL FIELD

Alexander L. Dmitriev

*St. Petersburg State University of Information Technologies, Mechanics and Optics Kronverkskiy Prospekt 49, St. Petersburg, 197101, Russia, alex@dmitriyev.ru*

**Abstract**. *On the basis of the field' concept of gravitation and gravitational analogue of the Faraday's induction law the difference of inertial mass of a body at its accelerated movement in horizontal and vertical directions relative to the Earth is shown. For an illustration of such a distinction the results of comparison of a motion of balance mechanical watch at horizontal and vertical orientations of balance' axis are given. The expediency of statement of precision mechanical experiments with measurement of anisotropy of the inertial mass is noted, allowing to estimate the validity of the "field" approach in the description of gravitation.*

**Keywords**: Gravitation, Acceleration, Inertial Mass, Anisotropy of Mass, Mach' Principle

**PACS**: 04.20.Cv, 04.80.-y

In distinction from "geometrical", the "field" concept of gravitation describes the gravitational interaction of bodies similarly to other kinds of physical interactions - electric and magnetic. Thus the concept of the "material" gravitational field related to sources - the gravitational mass - and characterized by the set of parameters (potential, velocity, impulse, moment) is considered. The advantage of the field, basically phenomenological concept of gravitation consists in an opportunity to use for its development some separate analogies of the gravitational and electromagnetic phenomena, and in their direct experimental check. Thus, gravitational fields, certainly, should have the properties similar, but not identical to properties of electromagnetic fields.

In [1] on the basis of the noted analogies the assumption of original reaction of the gravity force acting on a test body, on its acceleration $\vec{a}$ caused by action of external not gravitational (for example, elastic) forces is put forward. Change $\Delta \vec{g}$ of acceleration of the gravity, similar to the phenomenon of Faraday's induction law in view of Lenz' rule, is equal to

$$\Delta \vec{g}_{p,c} = -\alpha_{p,c} \vec{a}, \qquad (1)$$

where indexes $p, c$ indicate mutual, passing ($p$) or a contrary (opposite) ($c$), orientation of a vector $\vec{g}_0$ of normal acceleration of a gravity and vector $\vec{a}$ of acceleration of external force.

Estimations of the order of value of dimensionless factors $\alpha_p$ and $\alpha_c$, which the gravitational interrelation of gravitational and electromagnetic fields specify, were executed in mechanical experiments with weighing of two coupled mechanical rotors with the zero full mo-ment, with a horizontal axis of rotation, and in the analysis of the shock phenomena [2,3]. By consideration of thermal chaotic movement of microparticles of solid bodies the consequence 1, in view of an inequality $\alpha_p \succ \alpha_c$, is the negative temperature dependence of gravity, also observed in the experiment [4-7].

In [8,9] in the description of balance of the elastic (electromagnetic) and gravitational forces acting on the test mass on the part of remote mass (for example, stars), according to idea of E. Mach about the gravitational nature of inertial forces, the ratio between inertial ($m_i$) and gravitational ($m_g$) masses is obtained,

$$m_i = m_g (\alpha_p + \alpha_c). \qquad (2)$$

Equation 2 shows the direct proportionality of inertial and gravity masses of a body, and the relation of theses masses, contrary to the known postulate of "geometrical" model of gravitation, generally speaking, is not a constant.

Equation 1 shows the relation of change of gravity acceleration with acceleration $\vec{a}$ of external forces, in so doing it is necessary to take into account that the absolute size $\Delta g_{p,c}$ of an increment of acceleration should also depend on magnitude $g_0$ of normal gravity acceleration. Generally, in view of influence of forces of the gravitation caused by remote surrounding masses (stars), in movement of a test body on a vertical there should be carried out the equation

$$\alpha_{p,c} = A_{p,c} (g_0 + g'), \qquad (3)$$

where $g'$ - a projection of acceleration of forces of gravitation on the part of the remote masses located in a solid angle $2\pi$, on the direction of the accelerated movement of body, and factors $A_{p,c}$ characterize the action on a test body of not only Earth' gravitational field, but also a field of the gravitation created by all surrounding masses.

The resultant forces of gravitation acting on the motionless or moving with the constant speed test body from direction of remote masses, uniformly distributed in space in a full solid angle $4\pi$, it is approximately equal to zero, while the magnitude $g'$ determines the inertial properties of a body, Fig. 1.

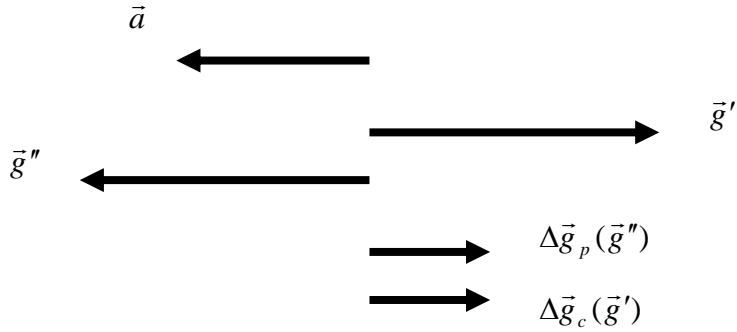

Fig. 1. Mutual orientation of a vector of acceleration of not gravitational forces $\vec{a}$ and increments vectors $\Delta\vec{g}_p, \Delta\vec{g}_c$ of accelerations of the gravitation forces acting on test mass from the direction of remote masses (stars). The resulting accelerations' vectors $\vec{g}'$ and $\vec{g}''$, caused by action of the remote masses located in the left and the right half-spaces in the solid angles $2\pi$, are equal in magnitude and are oppositely directed.

The consequence of 2,3 is the difference of inertial masses of a test body in its accelerated movement relatively to the Earth in horizontal and vertical directions.

For the harmonious, caused by the action of external elastic force, oscillatory movement along a vertical, the average, for the period of oscillation, inertial mass $m_i$ of a test body is equal to

$$m_i = m_g (A_p + A_c)(\frac{g_0}{2} + g'). \qquad (4)$$

In oscillatory movement of this test body along the horizontal, its average inertial mass $\overline{m}_i$ is equal to

$$\overline{m}_i = m_g (A_p + A_c) g'. \qquad (5)$$

In 4.5, the resulting magnitude $g'$ of projections of accelerations of gravity forces created by the remote masses in a solid angle $2\pi$, is believed constant and independent from the direction in space. The relative difference of "vertical" and "horizontal" inertial masses, taking $g' \succ\!\succ g_0$, is equal to

$$2\frac{m_i - \overline{m}_i}{m_i + \overline{m}_i} \approx \frac{g_0}{2g'}. \qquad (6)$$

Experimental estimations of magnitude of inertial mass anisotropy of a body can be made, comparing the periods of oscillations of linear mechanical oscillator with vertical and horizontal orientations of its axis. For the same purpose it is convenient to use the rotation oscillator, for example a pendulum of high-quality mechanical balance watch, by changing orientation of the balance axis.

The period $T$ of free oscillations of system a balance - spiral of mechanical watch is equal to

$$T = 2\pi \sqrt{\frac{I}{C}}, \qquad (7)$$

where $I$ - the moment of inertia of balance ($I \propto m_i$) and $C$ - factor of elasticity of the spiral [10].

According to 6,7, the period $T$ of oscillations of balance in a vertical plane should be more than the period $\bar{T}$ of oscillations of balance moving in a horizontal plane, that is the ideal mechanical watch in position "on an edge" goes more slowly, than in position "flatwise".

The position-sensitivity of mechanical watch is influenced many factors, including, the moment of inertia of a spiral, conformity of an axis of rotation and the centre of inertia of a pendulum, friction in axes of a suspension bracket of a pendulum etc [11]. With high quality of watch and its careful adjustment, the influence of the specified factors can be reduced practically to zero, and in that case the comparison of daily motion of balance watch in vertical and horizontal positions can be used for an estimation of magnitude of anisotropy of inertial mass 6. In view of 4-7, the relative difference $\gamma$ of the daily motion of an ideal watch is equal to

$$\gamma = 2\frac{T-\bar{T}}{T+\bar{T}} \approx \frac{1}{4}\frac{g_0}{g'}. \qquad (8)$$

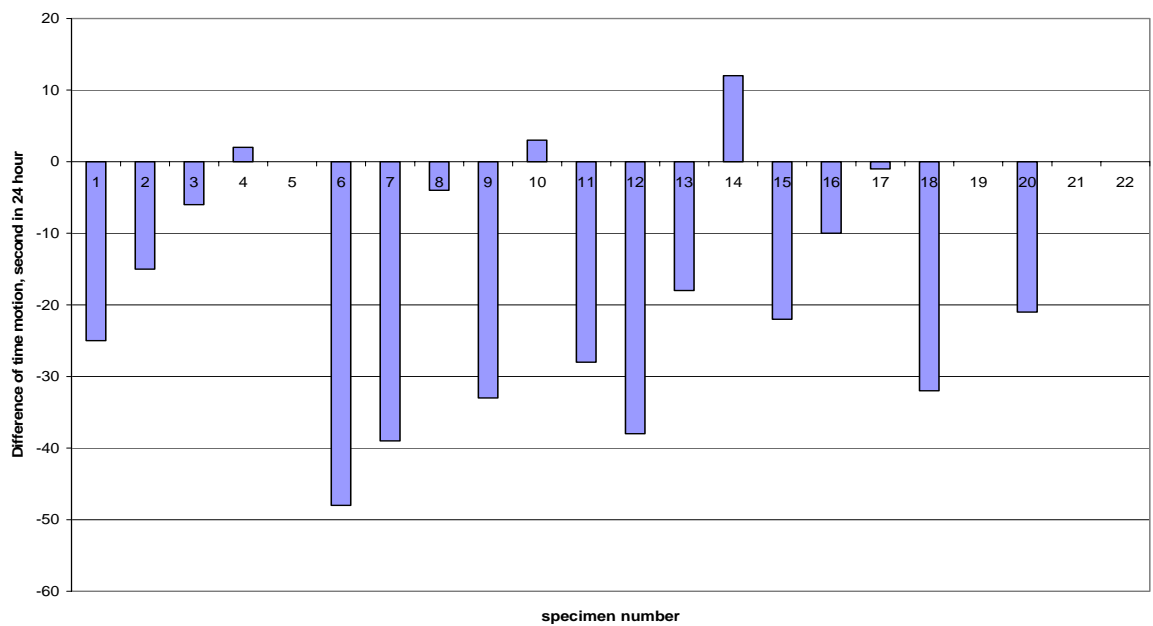

Fig. 2. A difference of a daily motion of mechanical balance watch "Raketa 2609" in positions "flatwise" and "on an edge".

In Fig. 2 the results of measurements of position sensitivity of twenty one samples of mechanical watch "Raketa 2609" manufactured by "Petrodvortsovy watch factory" are given. The difference of an average daily motion of watch in positions "flatwise" and "on an edge" was measured, each of them was measured as an average for two different positions of the head and plane of a dial of watch. The average magnitude of watch motion delay in position "on an edge" has come to about 15 seconds over one day which corresponds to $\gamma \approx 1.7 \cdot 10^{-4}$.

The question of what part of the given value $\gamma$ is caused by action of physical factors (anisotropy of inertial mass in a gravitational field of the Earth), and what – by technical imperfection of the mechanism of watch still remains open. The difficulty is that even with an appreciable influence on a motion of watch of anisotropy of inertial mass of the pendulum of watch, the position-dependence of a daily motion of watch can be reduced almost to zero by technical means of adjustment. Thus the "physical" delay of a watch motion can be compensated by adjustment of watch which complicates an objective estimation of magnitude of such effect. Therefore the careful analysis of all technical factors influencing the position sensitivity of balance watches and clockworks used in such experiments is necessary for obtaining of objective data. Nevertheless, the given average result is in agreement with physical preconditions noted above and can be the basis for setting up precision experiments with use of mechanical oscillators on measurements of prospective anisotropy of inertial mass.

For reduction of an error of the measurements connected to reorientation of an axis of rotation of mechanical pendulum, more reliable results, probably, will be received in measurements of self-frequency of linear mechanical oscillator with horizontally located axis which direction can be changed in a non-uniform gravitational field of the Earth (for example, in mountains).

If the result shown in Fig. 2 gives a true estimation of magnitude order of a relative difference of inertial masses in horizontal and vertical directions, then, according to 8, gravitational field-intensity $g'$ created by all indefinitely remote masses located in a solid angle $2\pi$ relative to a point of observation, the said intensity is approximately one thousand times the magnitude of normal acceleration of gravity on the surface of the Earth. In view of gravitational analogue of the Faraday's induction law 1, such rather strong "interstellar" gravitational field, apparently, is also the physical reason of inertial properties of bodies.

The precision measurements of anisotropy of inertial mass of bodies in a non-uniform gravitational field will confirm validity or an fallacy of the above estimation and as a consequence the validity of the phenomenological "field" concept of gravitation in the description of inertial properties of bodies.

**Appendix.**

If the average density $\rho$ of a matter in the volume of sphere of radius $R$ is constant, from the Newton's law of gravitation follows

$$g' = 2\pi G \rho R, \qquad (9)$$

where $G$ – the gravitational constant. Taking into account the temperature dependence of $G$ and non-uniform distribution of $\rho$, certainly, the 9 only for rough estimates is suitable.